\newcommand \be{\begin{equation}}
\newcommand \ba{\begin{eqnarray}}
\newcommand \ea{\end{eqnarray}}
\newcommand \ee{\end{equation}}

\documentclass[aps,prb,showkeys,
showpacs, 
floatfix,amsmath,
twocolumn,
10pt,
]
{revtex4-1}
\usepackage{epsfig}
\usepackage{color}
\usepackage{url}

\begin{document}

\title{Photon assisted L\'evy flights of minority carriers in $n$-InP}
\author{Oleg Semyonov}
\author{Arsen V. Subashiev}
\author{Zhichao Chen}
\author{Serge Luryi}
\affiliation{Department of Electrical and Computer Engineering,
State University of New York at Stony Brook, Stony Brook, NY,
11794-2350}
\begin{abstract}
We study the photoluminescence spectra of $n$-doped InP bulk wafers, both in the reflection and the transmission geometries relative to the excitation beam. From the observed spectra we estimate the spatial distribution of minority carriers allowing for the spectral filtering due to re-absorption of luminescence in the wafer. This distribution unambiguously demonstrates a non-exponential drop-off with distance from the excitation region. Such a behavior evidences an anomalous photon-assisted transport of minority carriers enhanced owing to the high quantum efficiency of emission. It is shown that the transport conforms very well to the so-called L\'evy-flights process corresponding to a peculiar random walk that does not reduce to diffusion. The index $\gamma$ of the L\'evy flights distribution is found to be in the range $\gamma=$ 0.64 to 0.79, depending on the doping. Thus, we propose the high-efficiency direct-gap semiconductors as a remarkable laboratory system for studying the anomalous transport.

\end{abstract}
\pacs{78.30.Fs,78.55.Cr,78.60.Lc,74.62.En}%
\keywords{photoluminescence, photon recycling, anomalous diffusion}

\maketitle 
\section{Introduction}
The shape of the luminescence spectra at room and cryogenic temperatures is known to be strongly 
affected by the doping: it modifies the absorption spectrum and therefore both the emission spectra and the radiation escape. More detailed physical picture of luminescence in heavily doped samples involves a number of different factors, such as change in the density of states, the bandgap narrowing, filling of the impurity and the conduction band states, carrier scattering and the resultant momentum non-conservation in optical transitions, as well as the effects of diffusion of the minority carriers followed by surface recombination. \cite{WILL} The relative impact of these factors is difficult to discriminate in heavily doped  samples ($N_D \ge 10^{18}$ cm$^{-3}$ ) and equally good fitting of the spectra can be obtained with different set of parameters (see Refs. \onlinecite{Sieg,Haufe,Rosen} ) leading to different  interpretations of the experimental data.

At moderate doping levels ($N_D \le 10^{18}$ cm$^{-3}$ ) the electron  system is non-degenerate and the luminescence spectra are not modified by the conduction band filling. The spectra remain similar for a whole range of moderate doping concentrations. The new phenomenon that comes into play in high-quality moderately-doped samples is the ``photon recycling'' that is the multiple radiative emission-reabsorption of luminescence photons.\cite{Dumke,Rossin,Roos} The radiative recombination processes in such samples are several orders of magnitude faster than the non-radiative recombination. Besides, the residual free-electron absorption of interband luminescent radiation becomes negligible. With negligible losses of minority carriers or photons, the kinetics of minority carrier luminescence is dominated by the ``photon recycling''.  Two main features of the recycling have been noted: \cite{Rossin} (i) the enhanced apparent lifetime of the minority carriers and (ii) enlarged minority carrier spread in the sample beyond the excitation area. Traditionally, the enhanced spread is interpreted in terms of a modified photon-assisted diffusivity and studied by numerical calculations,\cite{Badescu} which do not reveal the underlying physics. However, as will be shown in the present work, the spatial distribution of minority carriers is not exponential and the diffusion approximation fails completely.

In the photon-assisted diffusion model, the distribution of minority carriers is formed by two additive transport processes: (i) the random flights of holes (at sub-micron distances) interrupted by scattering, as in ordinary diffusion, and (ii) the photon-assisted transport of holes at much larger distances. The latter process corresponds to radiative recombination of a hole with the resultant interband photon, when reabsorbed, producing another hole at a different spot. If the reabsorption length is strictly limited, e.g. by the residual non-interband absorption, then the photon-assisted process can be viewed as a random walk with enlarged steps. On a large scale, this would be equivalent to a diffusive particle spread with much enhanced diffusivity. In this case, the hole distribution should decay exponentially away from the excitation area. The conventional hole diffusion mechanism implies a small diffusion length $ L_d \le 10$ $\mu$m, but with photon recycling one can envision at least a  tenfold enhanced $L_d$.  

However, in the experimental situation studied in this work, the effective $L_d \rightarrow \infty$ and the diffusion approximation fails as a matter of principle. The free path  of photons depends on their wavelength and even though the ``typical'' interband photons are re-absorbed at short distances, for those emitted in the red wing of the luminescence spectrum the free path grows exponentially and much exceeds the sample size. The overall distribution of the reabsorption probability and the distribution of the secondary holes created by reabsorption both acquire long non-exponential tails. The resultant random process of the temporal spread of the minority carrier concentration substantially differs from the normal diffusion and should be considered in terms of  L\'evy flights.\cite{Metz} The general L\'evy-flights (LF) transport corresponds to a random walk with the distribution of jump lengths ${\cal P} (x)$ characterized by a divergent second moment, e.g. ${\cal P}(x)\propto  |x|^{-(1+\gamma)}$ with the exponent $0 < \gamma < 2$.  The conventional diffusivity is not even defined for such a random walk. 

The LF transport problem has been extensively studied mathematically. Description of the anomalous transport in terms of fractional dynamical equations or, for random walks in an external field, fractional Fokker-Planck equations is amply discussed in the reviews.\cite{Metz,MetzKla} LF phenomena are well-known in astrophysics, as they occur in the problem of transport of resonance radiation in celestial bodies;\cite{Ivanov} they are also known in plasma physics as the imprisonment of resonance radiation in gaseous discharge.\cite{Holst,Pereira} Interestingly, these phenomena are more common than one would think: thus, L\'evy flights were recently invoked to explain movement strategies in mussels as revealed in the patterning of mussel beds,\cite{mussels} as well as ocean predators  search strategies in regions
where prey is sparse.\cite{Humph}


Here we report the results of  experimental and theoretical studies of the luminescence spectra of different $n$-type InP wafers. We demonstrate that these spectra are well suited to study the anomalous transport of the minority carriers.  The spatial distribution of minority carriers is revealed by the analysis of the  ratio of spectra observed in the transmission and reflection geometries. The observed spectra are distorted due to the energy-dependent re-absorption of outgoing radiation; we quantify this effect by filtering functions. Because of the exponential dependence of the absorption coefficient near the bandgap, the input of remote layers to the observed luminescence is strongly filtered and shifted to the red wing of the spectrum. The energy-dependent filtering  makes the experimentally observed spectra sensitive to the spatial distribution of minority carriers. 

The direct manifestation of the filtering effect is the red shift of the observed position of the maximum in the emission spectrum. This shift was clearly seen in the evolution of the luminescence line observed from the side edge of the sample when the distance from the excitation spot to the sample edge is varied up to several centimeters.\cite{SSCL} It was also manifested in the shape of the luminescence line observed in transmission geometry with thin samples.\cite{Semyon1}
  
In this paper, the minority carrier distribution in the samples is obtained by a Monte Carlo modeling of the photon recycling process. We assume that the intrinsic (unfiltered) emission lineshape is faithfully given by the van Roosbroek-Shockley (VRS) relation.\cite{VRSh} We then proceed to calculate the reabsorption probability in terms of the measured absorption coefficient and the VRS emission spectrum. The only remaining adjustable parameter is the quantum efficiency $\eta$ of emission and our experiments thus provide an accurate measurement of $\eta$. The quantum radiative efficiency can also be estimated independently from the time-resolved luminescence decay experiments\cite{Semyon1} and excellent agreement is obtained between these two determinations of $\eta$. The index $\gamma$ characterizing the L\'evy flights distribution is found to be dependent on the Urbach tailing parameter of the absorption spectrum and on the temperature. The observed reflection and transmission spectra for differently doped InP samples confirm the power-law decay of hole concentration and thus validate the concept of anomalous hole diffusion.  
\begin{figure*}
\begin{minipage}[b]{0.47\linewidth}
\centering\epsfig{figure=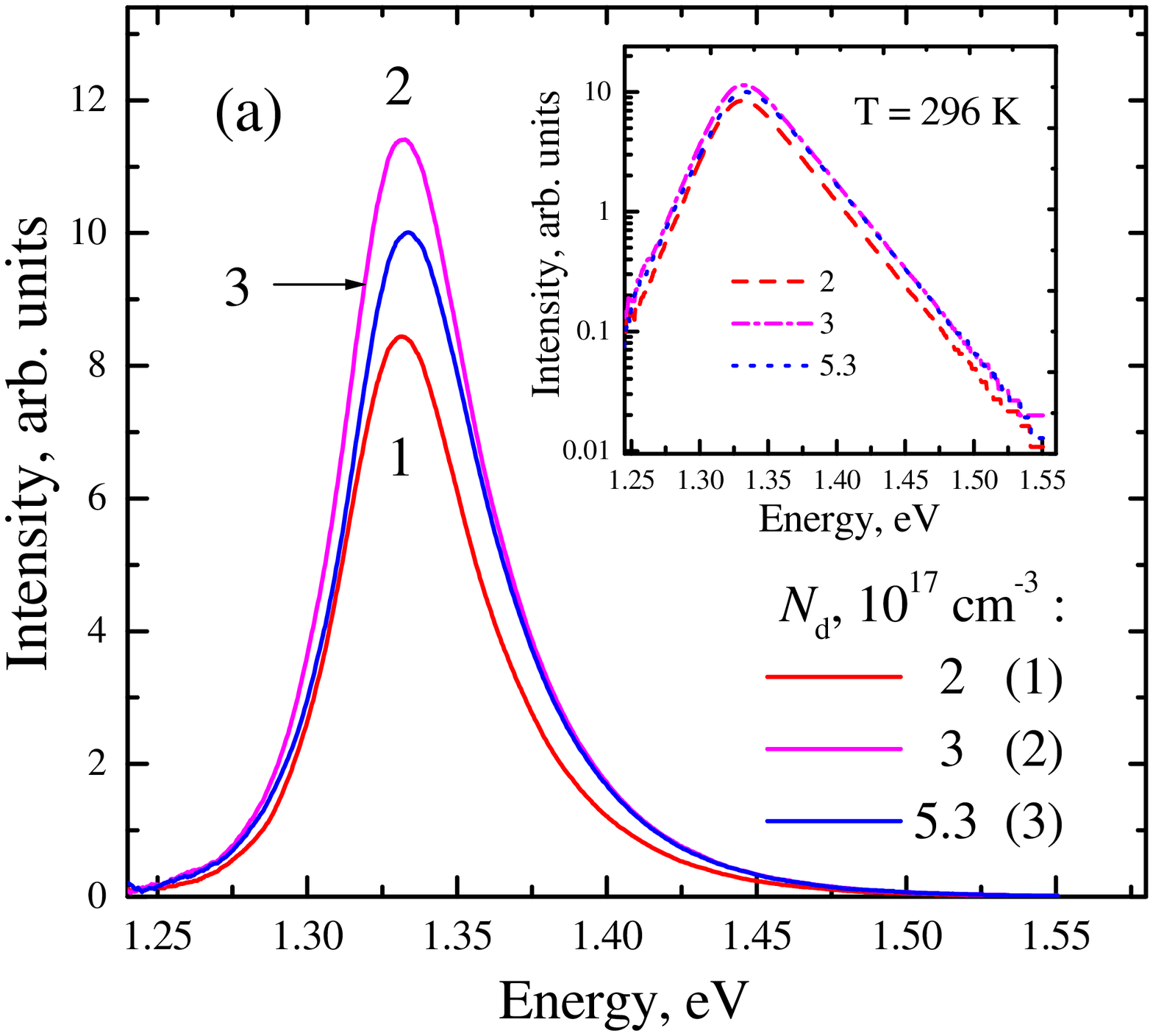,width=6.9cm}
\end{minipage}
\begin{minipage}[b]{0.51\linewidth}
\centering\epsfig{figure=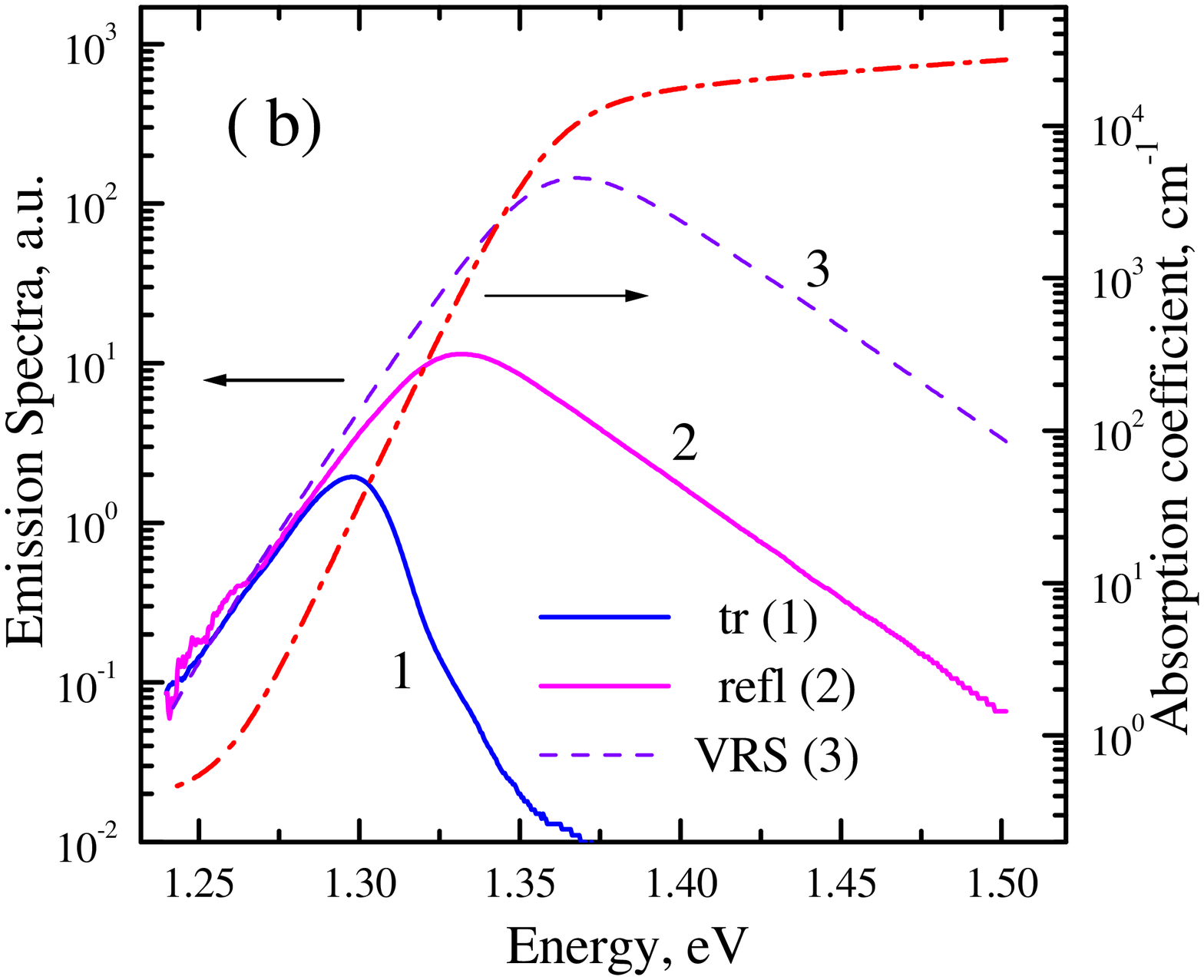,width=7.3cm}
\end{minipage} \hfill
\caption{(Color online) (a) Experimentally observed luminescence spectra in the reflection geometry for differently doped samples at T = 296 K. Inset shows the spectrum in the log scale. The right panel (b) shows the luminescence spectra both  in reflection and transmission geometries for $N_D=3\times 10^{17}$ cm$^{-3}$. The dashed line (VRS) shows the intrinsic emission spectrum calculated with the van Roosbroek-Shockley relation, Eq. (\ref{ShVR}). The dash-dotted line shows the absorption coefficient for this sample.}
\label{SpecvsX}
\end{figure*}
\section{Experimental results and their interpretation}
\subsection*{Experimental spectra}
The luminescence spectra were measured in moderately-doped ($N_D \le 10^{18}$ cm$^{-3}$ ) and heavily-doped (up to $N_D = 8\times 10^{18}$ cm$^{-3}$ ) InP wafers of thickness $d$= 350 $\mu$m  supplied by Nikko Metals.\cite{Acro} For some doping levels thinned samples were also studied. The spectra were taken  at room temperature in both reflection and transmission geometries. The excitation wavelength $\lambda = 640$ nm  was chosen  to ensure short penetration of the incident radiation into the wafer, so that the resulting distribution of holes is dominated  by the carrier kinetics. The distribution is slightly modified only near the surface in a region whose thickness depends mainly on the surface recombination rate. Luminescence spectra observed in the reflection geometry for moderately doped samples with $N_D=2,3$ and $5.3\times 10^{17}$ cm$^{-3}$ are presented in Fig. \ref{SpecvsX} (a). The inset shows the same spectra in the logarithmic scale, demonstrating an exponential drop-off on both red and blue sides. For all moderately doped samples  the observed spectra are similar, varying only in the luminescence intensity. The spectra in both reflection and transmission geometries are shown in log scale in Fig. \ref{SpecvsX} (b) for a sample with   $N_D=3\times 10^{17}$ cm$^{-3}$. We have also measured the absorption curves in the transparent region up to the values $\alpha(E) \approx 200$ cm$^{-1}$ limited by the transparency of the samples. To obtain the absorption spectra in a broader energy region, we have extrapolated our experimental data using the available data obtained with much thinner samples. \cite{Semyon1,Augustin,textbook} The entire absorption spectrum is shown in Fig. \ref{SpecvsX} (b) by a dash-dotted line; also shown is the intrinsic emission spectrum $S_{VRS}$, calculated with the van Roosbroeck-Shockley relation,\cite{VRSh,cube}
\be
S_{VRS}(E) \propto  E^3 \alpha_i(E)\exp(-E/kT) \label{ShVR}~.
\ee
Here $\alpha_i(E)$ is the interband contribution to the absorption coefficient $\alpha(E)$ calculated as $\alpha_i(E)=\alpha(E)-\alpha_{fc}$, where $\alpha_{fc}$ is the free-carrier absorption coefficient (associated with the interband transitions to the upper conduction valley\cite{dumkeFCA}). The $\alpha_{fc}$ is linear in doping and does not tangibly vary with energy in the region of interest. Therefore, its subtraction can be done with sufficient accuracy.\cite{SSCL}

Figure \ref{SpecvsX} (b) clearly indicates the red shift of the observed luminescence spectra relative to the intrinsic emission spectrum and the absorption spectrum of the sample. Both in the transmission and reflection geometries the maxima of observed lines are on an exponentially steep slope of the absorption spectrum. The reflection-geometry spectra at low doping (up to $5\times 10^{17}$ cm$^{-3}$) are shifted to the red by about 30 meV as compared to the VRS line.  Spectra in the transmission geometry show a much larger red shift (of $\approx 70$ meV) and a steeper decay at the high-energy side.  The distortion of the intrinsic VRS spectrum is due to reabsorption of the outgoing radiation. 
\begin{figure}[b]
\epsfig{figure=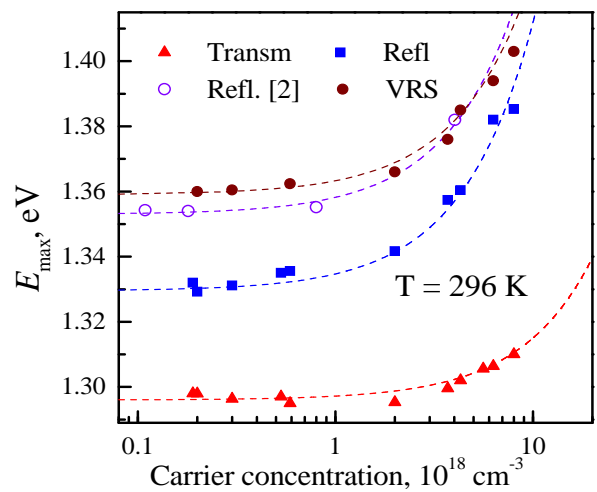,width=7.8cm,height=6.5cm} \caption[]
{(Color online) Positions of the maxima of the photoluminescence spectra in reflection and transmission geometries for differently doped samples. Open dots show the results in reflection geometry from epitaxial samples by Sieg {\it et al.} \cite{Sieg} The maxima calculated for van Roosbroek-Shockley (VRS) spectra are shown by full dots. Dashed lines are guides for the eye.}
\label{Maxima}
\end{figure}

The effects of reabsorption can also be traced in the shifts of the peak energies $E_{\rm max}$ of the photoluminescence spectra relative to those of the intrinsic spectra. These shifts in both reflection and transmission geometries are illustrated in Fig. \ref{Maxima} for moderately-doped and heavily-doped samples. Shown are the $E_{\rm max}$  values obtained in this work for differently dopes samples with $d=350$ $\mu$m as well as those observed by Sieg {\em et al.}\cite{Sieg} in reflection geometry for epitaxial layers of thickness 2 $\mu$m. The epitaxial results exhibit much smaller reabsorption red shifts. For comparison, Fig. \ref{Maxima} also shows the  $E_{\rm max}$ positions  for  intrinsic spectra calculated with Eq. (\ref{ShVR}). The  steady increase of $E_{\rm max}$ with the majority carrier concentration is explained by the conduction band filling and the Moss-Burstein shift. The smaller increase of $E_{\rm max}$  at high doping for the transmission spectra is due to the smaller spread of the minority carriers, as discussed below.    
\begin{figure*}[t]
\begin{minipage}{0.48\linewidth}
\epsfig{figure=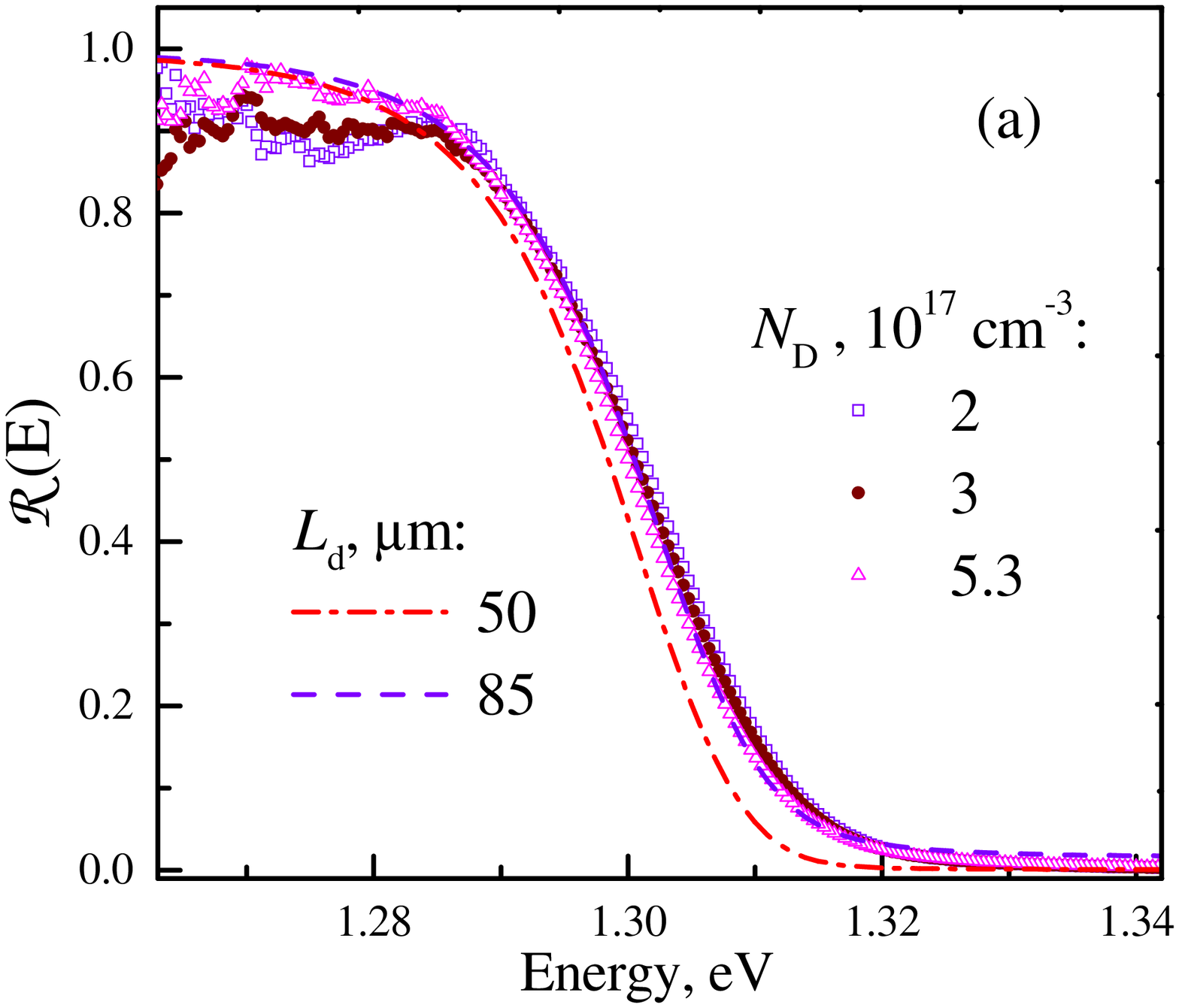,width=7.2cm,height=6.7cm} \end{minipage} \hfill
\begin{minipage}{0.50\linewidth}
\centering\epsfig{figure=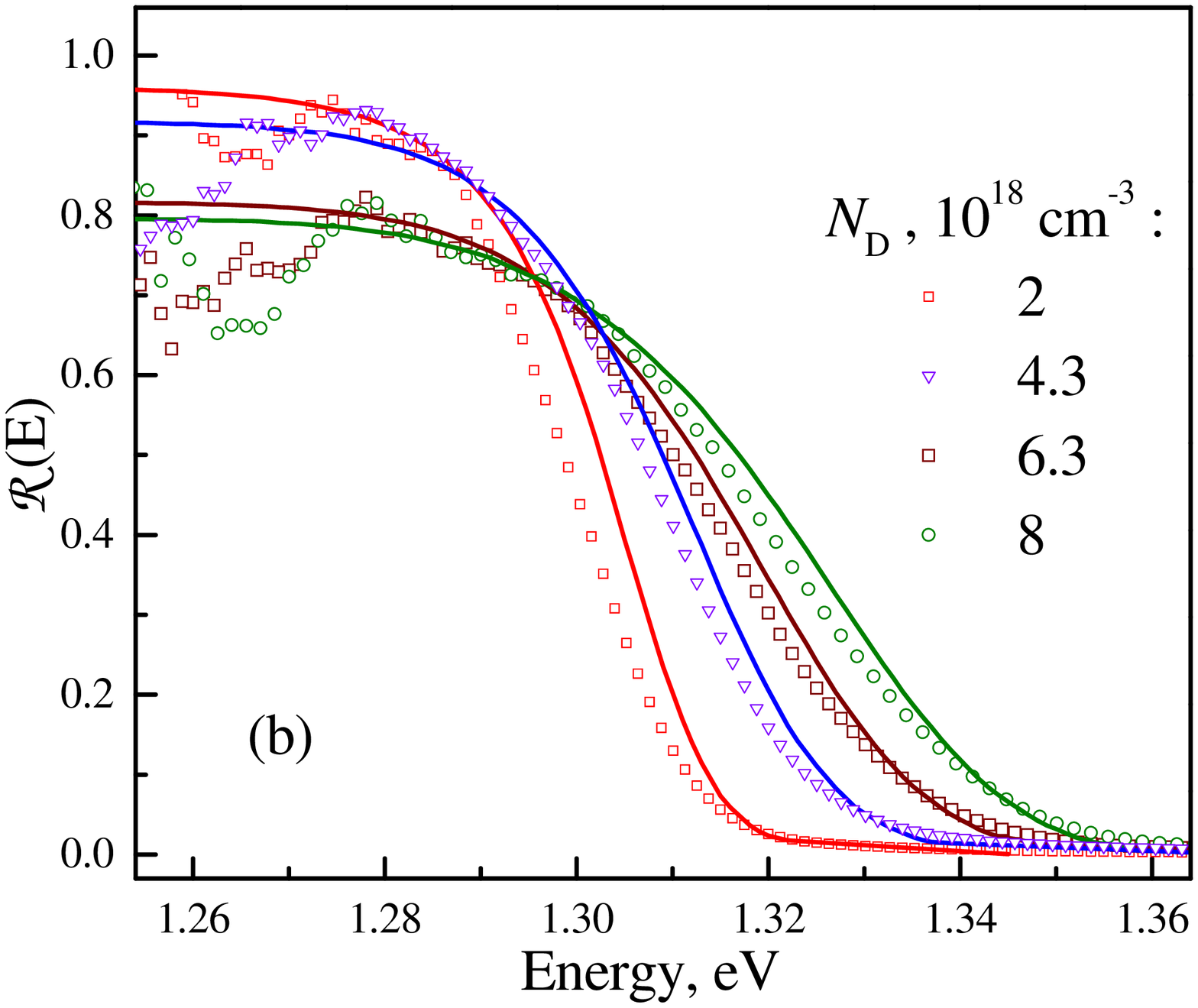,width=7.2cm,height=6.7cm}
\end{minipage} 
\caption[]{(Color online)  (a) Ratio of the spectra observed in transmission and reflection geometries, $\mathcal{R}(E) = I_{tr}(E)/I_{refl}(E)$ for moderately doped samples (dots), $d=350$ $\mu$m. Lines show the ratio $\mathcal{R}(E)$ calculated for a sample with  $N_D=3\times 10^{17}$ cm$^{-3}$, assuming exponential distributions of the minority carriers corresponding to diffusion lengths $L_d=$ 50 $\mu$m and 85 $\mu$m. (b) Ratio  $\mathcal{R}(E)$ for  heavily-doped samples (dots). Solid lines show the results of calculations based on Eqs. (\ref{refFil},\ref{trFil}) using the diffusion approximation for the hole distribution with diffusion length as a fitting parameter, see Table \ref{tab:Ld}.}
\label{FiltRatio}
\end{figure*}

\subsection*{Interpretation: filtering functions}
The observed spectrum in reflection geometry can be described by the following expression
\be
I_{refl}(E)= S_{VRS}(E)F_{refl}(E)~, \label{SpecR} 
\ee
which defines the filtering (reabsorption) function $F_{refl} (E)$. 
The spectrum observed in transmission geometry can be similarly expressed in the form
\be
I_{tr}(E)= S_{VRS}(E)F_{tr}(E)~, \label{SpecT} 
\ee
where $F_{tr} (E)$ is the filtering (reabsorption) factor in transmission geometry. 

The ratio of experimental spectra, $\mathcal{R}(E)=I_{tr}(E)/I_{refl}(E)$, measured in transmission and reflection geometries is shown  by the dots in Fig. \ref{FiltRatio} (a) for three moderately doped samples with different doping levels. According to Eqs. (\ref{SpecR},\ref{SpecT}), this ratio reduces to the ratio of the corresponding filtering functions, $\mathcal{R}(E)=F_{tr}/F_{refl}$, with the factor $S_{VRS}$ canceling out. The ratio $\mathcal{R}(E)$ is convenient for our analysis becasue it does not depend on the details of the iintrinsic VRS spectrum. For all samples a steep decline of the $\mathcal{R}(E)$ with $E$ is observed in the region from $E\approx 1.27$ eV to $E\approx 1.31$ eV, with a minor shift of the curve to lower energies for higher doped samples. 

The absorption and luminescence spectra for heavily doped samples were reported earlier.\cite{Semyon1}. These spectra strongly depend on the doping, predominantly due to  the conduction band filling and the attendant Moss-Burstein effect. For heavy doping, the luminescence lineshapes are poorly described by the VRS relation. Presumably, while in low doped samples the shape of the spectra is dominated by the Urbach exponential decay in the red wing and the temperature-dependent exponential decay in the blue wing, the main input for the heavily-doped samples comes from the absorption region $\alpha = 10^3 - 10^4$ cm$^{-1}$ where the behavior $\alpha (E)$ is non-exponential. Measurements  of the energy dependence in this region have a large uncertainty.   Experimental results for the ratio $\mathcal{R}(E)$ for these samples are shown by dots in Fig. \ref{FiltRatio} (b). In this case the ratio variation depends on the doping. The Moss-Burstein shift manifests itself in the higher-doped samples as a shift in the curves $\mathcal{R}(E)$ to the high-energy side. 
\begin{figure}[b]
\epsfig{figure=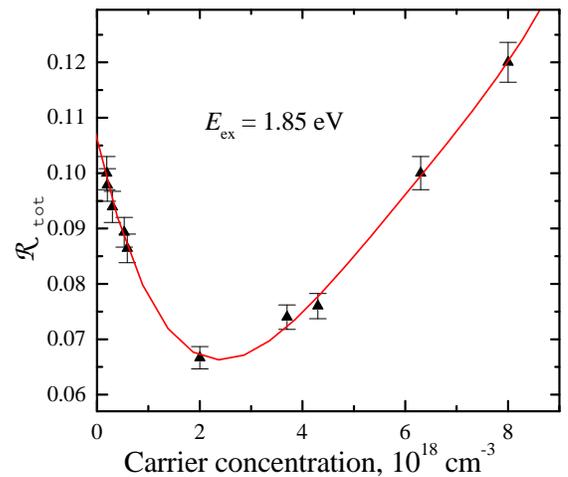,width=7.3cm} \caption[]
{(Color online) Ratio of the total luminescence intensities $\mathcal{R}_{tot}$ for differently doped samples, $d = $ 350 $\mu$m. Solid line is a guide to the eye.}
\label{TotIntenR}
\end{figure}

\begin{figure*}[t]
\begin{minipage}{0.46\linewidth}
\centering\epsfig{figure=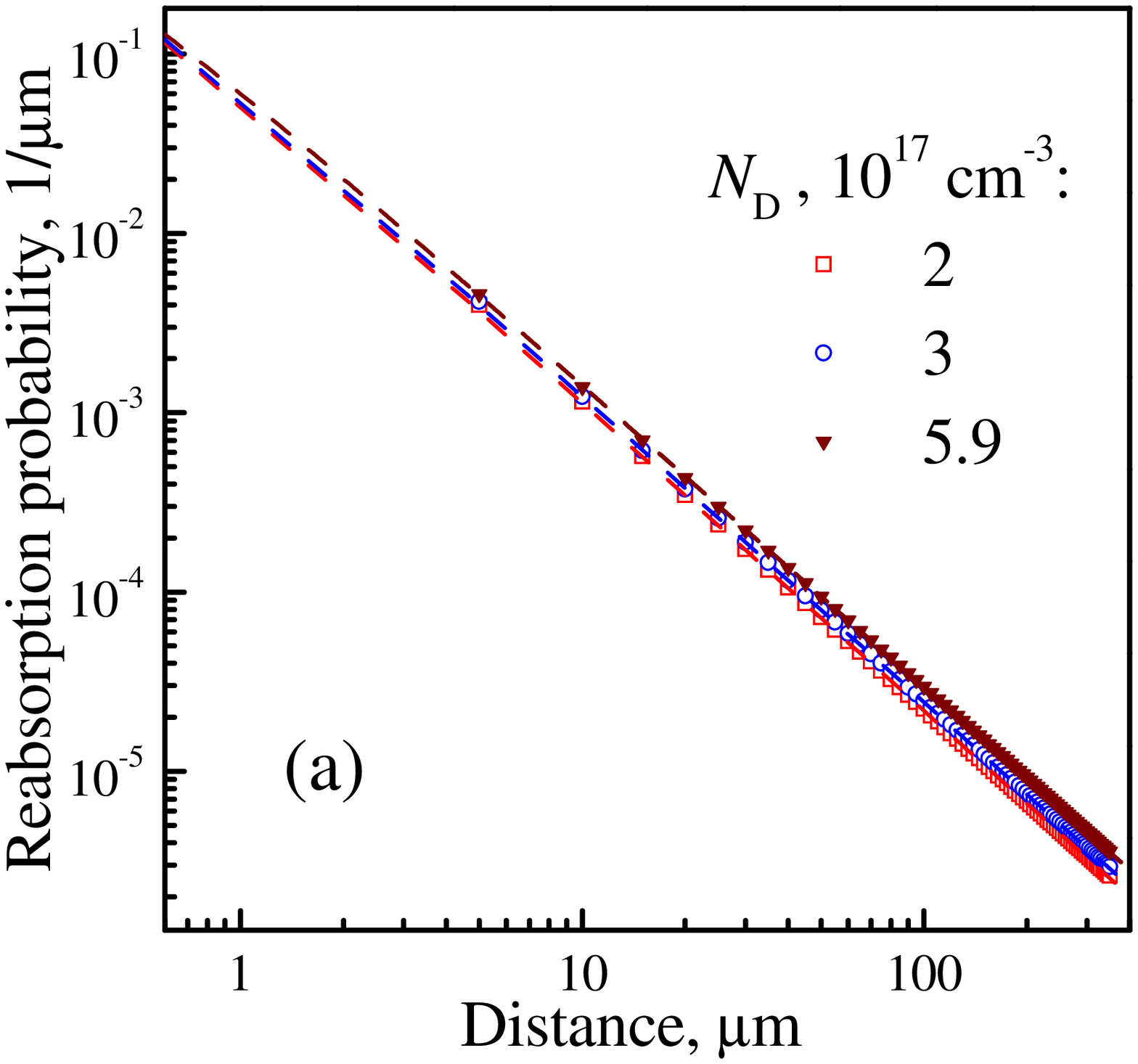,width=6.7cm}
\end{minipage}
\begin{minipage}{0.50\linewidth}
\centering\epsfig{figure=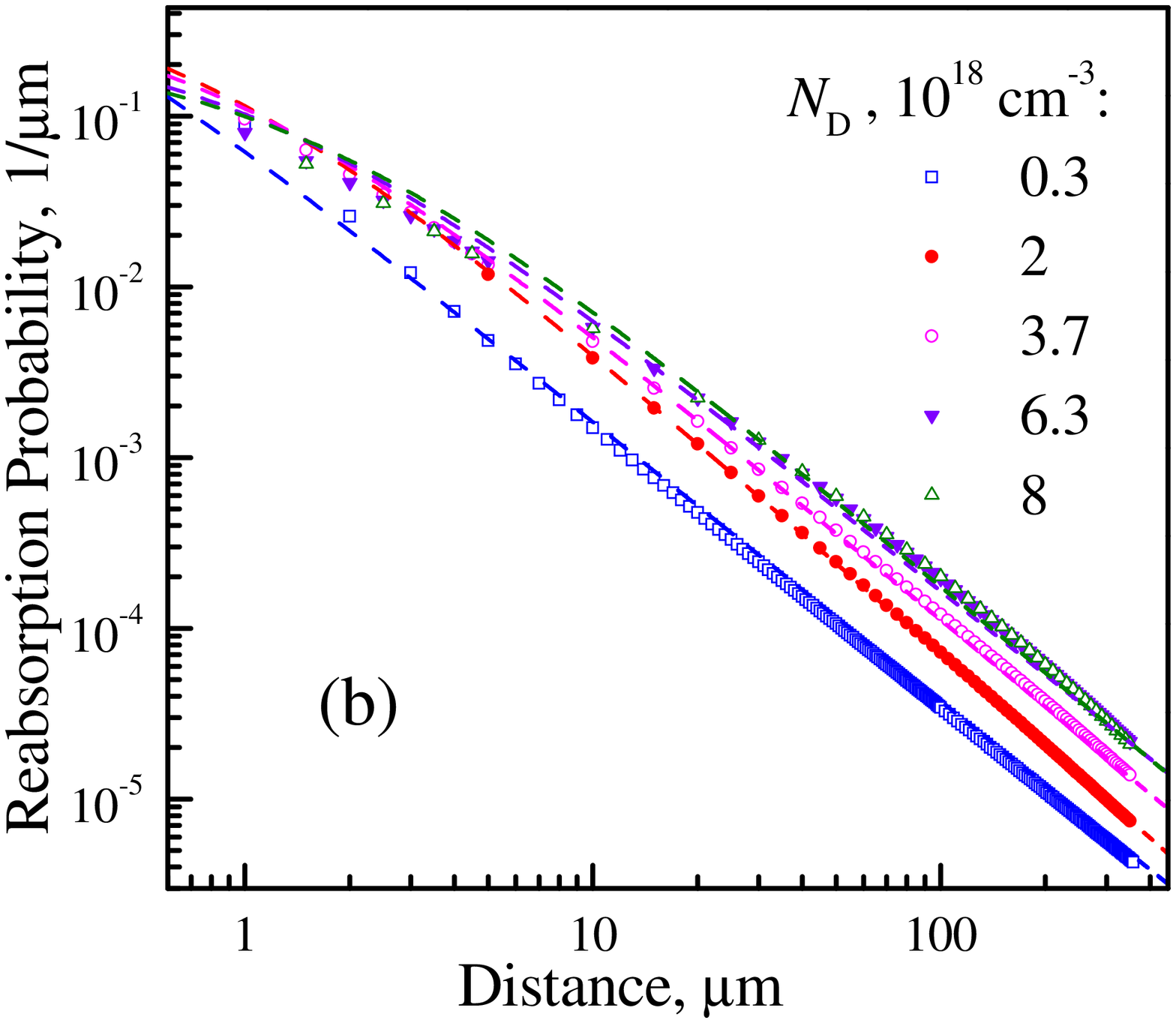,width=6.9cm}
\end{minipage} \hfill
\caption{(Color online) Reabsorption probability ${\cal P}(z)$ for moderately doped samples (a) and for heavily doped samples (b) calculated with Eq. (\ref{Pro}) (dots). Dashed lines correspond to the power-law approximation, Eq. (\ref{probKer}).}
\label{KernLDe18}
\end{figure*}

The effect of the Moss-Burstein shift is clearly seen in experimental results for the ratio of the total (integrated) intensities of the emission lines $\mathcal{R}_{tot}=I_{tot,tr}/I_{tot,refl}$ for the whole set of differently doped samples presented in Fig. \ref{TotIntenR} as a function of the doping level, $\mathcal{R}_{tot}(N_d)$. Two notable features of this dependence are: (i) high values of the transmitted radiation intensity; and (ii) minimum of $\mathcal{R}_{tot}$ at $N_d \approx 2\times 10^{18}$ cm$^{-3}$. The latter effect reflects subtle details of the reabsorption probability (cf. Fig. \ref{KernLDe18}) and will be discussed below.  

To check the influence on the spectra of the in-plane distribution of the excitation, we varied both the size of the excitation spot and the size of the  registration area of the observed spectra. Only minor changes in the spectra were observed. This indicates that the only dependence of importance is the distance from the  surface of excitation. Hence one-dimensional kinetics model is adequate for the spectra interpretation. 

It is important that due to the high refractive index of InP, the total internal reflection angle for radiation coming out into air is $\theta_t$=17.5$^\circ$ [so that $ \cos(\theta_t) = 0.954$]. As a result, the spread of the travel distances for outgoing radiation is negligible. We assume therefore that the observed radiation propagates along the normal to the sample surface (we have checked that allowing for the angular spread does not lead to sizable corrections).  The filtering factors $F_{refl}(E)$ and $F_{tr}(E)$ can be expressed through the one-pass filtering functions $F_1(E)$ and $F_2(E)$, defined by
\ba 
F_1(E)&=& \int_0^d p(z)\exp[-\alpha(E)z] dz,  \nonumber \\
F_2(E)&=& \int_0^d p(z)\exp[\alpha(E)(d- z)] dz~,  \label{Filt12}
\ea
where $p(z)$ is the non-equilibrium hole concentration at a distance $z$ from the sample surface.

For a sample  of finite thickness, taking into account multiple reflections of the luminescence radiation, we have
\be  
F_{refl}(E)=(1-R)\frac{F_1+R F_2 \exp(-\alpha d)}{1-R^2  \exp(-2\alpha d)}~, \label{refFil}
\ee
and
\be
F_{tr}(E)=(1-R)\frac{F_2+R  F_1\exp(-\alpha d)}{1-R^2  \exp(-2\alpha d)}~.  \label{trFil}
\ee
Here $R$ is the InP reflection coefficient. For our samples it was measured to be $R=0.3$ close to the absorption edge; in the further numerical calculations we took into account the experimental dependence $R(E)$ and its variation with doping.

As seen from Eqs. (\ref{SpecR},\ref{SpecT}) and (\ref{refFil},\ref{trFil}), the ratio
\be \mathcal{R}(E)=\frac {F_{tr}(E)} {F_{refl}(E)}=\frac {F_2+R  F_1\exp(-\alpha d)}{F_1+R F_2 \exp(-\alpha d)}  \ee 
has important advantages for the analysis of the minority carrier profile. As noted already, this ratio does not depend on details of the intrinsic emission spectrum. Furthermore, it is not sensitive to multiple reflections, since the denominators of Eqs. (\ref{refFil},\ref{trFil}) cancel out. At the same time, it is quite sensitive to $p(z)$ through $F_1(E)$ and $F_2(E)$. Therefore, the ratio $\mathcal{R}(E)$ can be used to quantify the spatial hole distribution. 

For a rough estimation of the hole spread, we have calculated the ratio $\mathcal{R}(E)$  assuming that the hole concentration decays exponentially away from the surface, $p(z)=p(0) \exp(-z/L_d)$, where $L_d$ is an ``effective diffusion length'' that we use as a fitting parameter. The results  for moderately doped samples are shown in Fig. \ref{FiltRatio} (a) by dashed lines for two values of $L_d$. We see that  $L_d=85$ $\mu$m describes the experimental results very well, while the smaller value $L_d=50$ $\mu$m gives a steeper decay than that observed experimentally. Both values are far above the typical hole diffusion lengths (which do not exceed 10 $\mu$m). 

For the heavily doped samples, Fig. \ref{FiltRatio} (b),  the exponential approximation for $p(z)$ does not fit the experimental data quite well. The best fit procedure gives the values of $L_d$ listed in Table \ref{tab:Ld}. The energy variation of the ratio is dominated by the strong decay of $F_{tr}(E)\propto \exp[-\alpha(E) (d-L_d)]$, while the decrease of $F_{refl}(E)$ is shifted to higher energies due to the shorter escape distance. Therefore, any distribution with an average spread length $L_d$ that is considerably shorter than the layer thickness $d$ will produce curves similar to those depicted in Figs. \ref{FiltRatio}.  As seen from Table \ref{tab:Ld}, the hole spread in the layer is reduced in heavily doped samples but still is much broader than could be anticipated for a conventional hole diffusion, indicating the importance of recycling effects. 
\begin{table}
\caption{\label{tab:Ld} Effective diffusion length for differently doped samples obtained from the data for thick $d=350$ $\mu$m samples.}
\begin{ruledtabular}
\begin{tabular}{llllll}
  $N_D$, 10$^{18}$ cm$^{-3}$ & 0.2-0.6& 2 & 4.3 & 6.3 & 8 \\
  $L_d$, $\mu$m & 85 & 45 & 43 & 40 & 10\\
\end{tabular}
\end{ruledtabular}
\end{table}

Note that the exponential decay approximation with an enhanced effective diffusion length fails to describe the variation of $\mathcal{R}(E)$ with the sample thickness, especially for thin ($\approx 50$ $\mu$m) samples.

As a matter of principle, accurate measurements of the reflection and transmission spectra together with the absorption spectra enable us to find $F_{refl}$ and $F_{tr}$ and from these functions recover $p(z)$. Indeed, using the experimentally known function $\alpha(E)$  one can calculate $S_{VRS}(E)$. As discussed above, this step works well for moderately doped samples. Next, we find  $F_{refl}(E(\alpha))$ and $F_{tr}(E(\alpha))$ via Eqs. (\ref{SpecR},\ref{SpecT}). From these functions one can find the one-pass filtering functions (\ref{Filt12}), regarding them as functions of $\alpha$, viz.
\ba
F_1(\alpha)&=&\frac{F_{refl}-R F_{tr} \exp(-\alpha d)}{1-R}~, \nonumber \\
    F_2(\alpha)&=&\frac{F_{tr}-R F_{refl} \exp(-\alpha d) }{1-R}.    \label{FilFunk}
\ea

Equations (\ref{FilFunk}) with their right-hand sides expressed in terms of the experimental data represent experimental determination of the one-pass filtering functions. Then Eqs. (\ref{Filt12}) can be viewed as integral equations for $p(z)$. In fact, they represent a Laplace transformation (with the argument $\alpha$) of the hole distribution function $p_r(z)$, defined (for $F_1$) as $p_r(z)=p(z)$  for $0<z<d$ and $p_r(z)=0$ otherwise. Similarly, for $F_2$ one can consider $p_t(z)=p (d-z)$ inside the sample. In principle, one can obtain $p(z)$ by inverting the Laplace transformations numerically, using one of the developed approaches.\cite{DavMar} The pair of Eqs. (\ref{FilFunk}) can be used for a subsequent reiteration scheme in the numerical solution. However, the numerical inversion of the Laplace transformation is a classical ill-conditioned mathematical problem.\cite{Num} Attempts to find  $p(z)$ in this way suffer from a low solution accuracy. The dominating errors in the filtering functions appear in the red-wing region of the experimental spectra, where the luminescence intensity is low and the noise signal is high. This high-noise signal in the low-energy region (where $F_1 \approx F_2 \approx 1$) is clearly seen in Fig. \ref{FiltRatio}. 

Therefore, instead of attempting to invert the Laplace transformations, we have modeled the distribution of holes in the layer, calculated the filtering functions, and compared the results with the functions found from the experiments. 

\section{Minority carrier distribution in a finite layer}
Studies of the temporal relaxation of non-equilibrium minority-carrier excitations in doped $n$-InP demonstrate a very high quantum radiation efficiency and high values of the photon recycling factor\cite{Semyon1} (which in the bulk  crystal equals the ratio of radiative to non-radiative recombination rates). Therefore the hole distribution in case of the stationary excitation is strongly influenced by the recycling process. In an infinite medium with the initial excitation homogeneously distributed in the plane $z=z'$  the probability of reabsorption in the plane $z=z''$ is given by \cite{Rossin,Ivanov}
\be
{\cal P}(z''-z')= \frac 1 {2} \int {\cal N}(E) \alpha_i(E){\rm Ei}[1,\alpha(E)|z''-z'|] dE~. \label{Pro}
\ee
Here  ${\cal  N}(E)$ is the normalized spectral density for the number of photons in intrinsic luminescence, ${\cal N}(E)=CS_{VRS}(E)/E$, cf. Eq. (\ref{ShVR}), and $C$ is a normalization constant. The exponential integral function Ei$(1,|z|)$ is defined by
\be
{\rm Ei}(1,z)=\int_1^\infty \frac {dt}{t}\exp(-zt)~. \label{expint}
\ee
Due to the residual absorption the full probability of  interband reabsorption $P_{tot}<1$.  However, in the range of thicknesses and distances of interest the difference is negligible for all our samples, so that
\be
{\cal P}_{tot}=\int_{-\infty}^{\infty} {\cal P}(z) dz =  \int_{0}^{\infty} {\cal N}(E) \frac {\alpha_i(E)}{\alpha(E)}dE  \approx 1 ~. \label{Norm}
\ee
Results of numerical evaluations of ${\cal P}(|z|)$  with Eq. (\ref{Pro}) are shown in Fig. \ref{KernLDe18} (a). For the entire range of $z$, the dependence ${\cal P}(|z|)$  is very close  to
\be
{\cal P}(|z|)=\frac {\gamma z_{min}^{\gamma}}{2(z_{min}+z)^{1+\gamma}}~. \label{probKer}
\ee
For moderately doped samples, $z_{min}\approx$ 0.1 $\mu$m and the index $\gamma\approx 0.65$ while slightly decreasing with the doping level.
\begin{table}
\caption{\label{tab:Kern} Parameters of the reabsorption probability distribution (\ref{probKer}) and the recycling factor $\Phi$  for differently doped samples; the $\Phi_l$ values are from Ref. \onlinecite{Semyon1}.}
\begin{ruledtabular}
\begin{tabular}{llllll}
  $N_D$, 10$^{18}$ cm$^{-3}$ & 0.2-0.6  & 2     & 3.7   & 6.3    & 8 \\
  $\gamma$            & 0.64          & 0.79  & 0.7   & 0.64   & 0.69\\
  $z_{min}$, $\mu$m    & 0.0962        & 0.625 & 0.7   & 1     & 1.377\\
  $\Phi$    & 90-97    & 32  & 18  & 10 & 8  \\
  $\Phi_l$    & -  & 48 & 20 & 12 & 9 \\
\end{tabular}
\end{ruledtabular}
\end{table}
\begin{figure*}[t]
\begin{minipage}{0.46\linewidth}
\centering\epsfig{figure=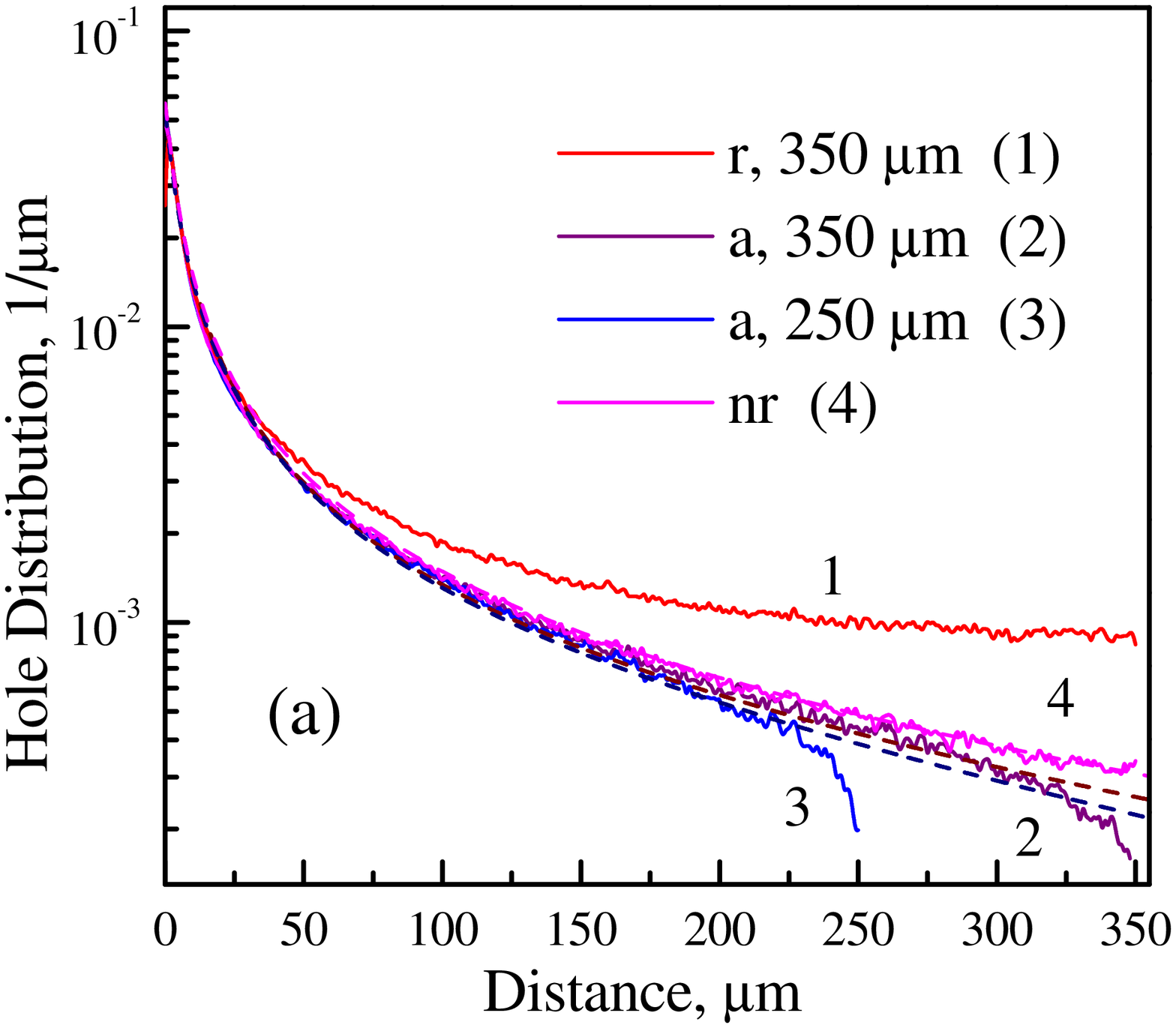,width=6.9cm}
\end{minipage}
\begin{minipage}{0.50\linewidth}
\centering\epsfig{figure=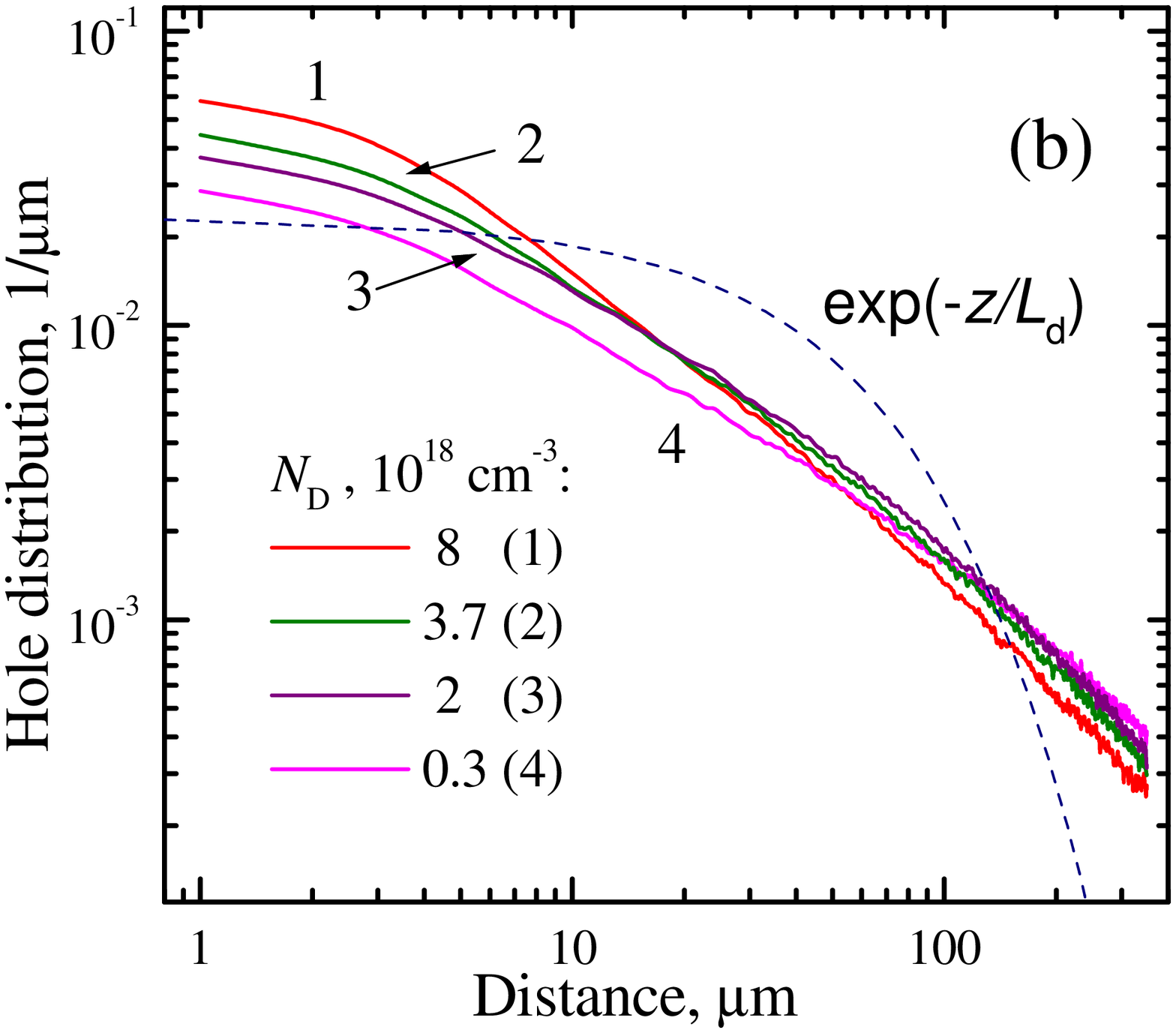,width=6.9cm}
\end{minipage} \hfill
\caption[]
{(Color online) (a) Stationary hole distribution $p(z)$ (in log-linear scale)  calculated by Monte Carlo with the reabsorption probability (\ref{probKer}) for  $N_D=6.3 \times 10^{18}$  cm$^{-3}$ and different boundary conditions at the back surface  (r: reflective, a: absorptive, and nr: non-reflective). Dashed lines show the fitting with the power law $p(z)=c/(z+z_{min})^{1+\tilde \gamma}$. (b) The hole distribution (in log-log scale) for differently doped samples; the sample parameters are listed in Table \ref{tab:Kern}. The dashed line corresponds to an exponential distribution with $L_d=45$ $\mu$m (cf. Table \ref{tab:Ld}). }
\label{HoleDist}
\end{figure*}

The emission-reabsorption process described by Eq. (\ref{Pro}) can be interpreted as a one-dimensional random walk of holes. Its stochastic nature is associated with the emission spectrum ${\cal  N}(E)$. The flight probability, accurately described by Eq. (\ref{probKer}), is typical for the ``anomalous diffusion'' of the L\'evy-flights type. Its hallmark is the asymptotic spatial decay ${\cal P}(z) \propto z^{-1-\gamma}$ with a ``heavy tail'', i.e. the power-law asymptotics with an index $\gamma <2$. Moments of this distribution  diverge, and so does the conventionally defined diffusion coefficient. 

Analytical calculation of the index is possible in a model where the interband absorption spectrum is approximated by a simple function 
\be
\alpha_i(E) = \frac{\alpha_0}{1+\exp[(E_g-E)/\Delta]}~,
\label{TailAbso}
\ee
which decays exponentially below the absorption edge and saturates above it. This model accounts for the Urbach tailing (described by the tailing parameter $\Delta$), but it does not describe the growth of $\alpha_i(E)$ at $E\ge E_g$. Similarly, the emission spectrum can be approximated by the VRS relation (\ref{ShVR}) [with  $\alpha_i(E)$ given by Eq. (\ref{TailAbso})] in which the pre-exponential factor  for $\Delta \le kT \ll E_g$ can be replaced by its value at $E_g$. In this model one obtains
\be
\gamma = 1-\frac{\Delta}{kT} ~.
\label{indexM}
\ee

Equation (\ref{indexM}) predicts a decrease of $\gamma$ at lower temperatures. It also explains the decrease of $\gamma$ with increasing $N_D$. The later effect is due to the smearing of the absorption edge (described by an increase of the Urbach tailing parameter $\Delta$).  Estimation of $\gamma $ with Eq. (\ref{indexM}) for moderately doped samples ($\Delta = 9.4$ meV) \cite{SSCL} gives $\gamma \approx 0.64$, close to the results obtained with more accurate numerical calculations.   


For heavily doped samples, the distributions ${\cal P}(|z|)$ calculated numerically with Eq. (\ref{Pro}) are shown in  Fig. \ref{KernLDe18} (b) with the distribution parameters listed in Table \ref{tab:Kern}. The  calculated variation of $\gamma$ is more complicated than that predicted by Eq. (\ref{indexM}) due to the competition of several factors. Thus, compared to the low-doped samples, the index increases to $\gamma=0.78$ for $N_D=2\times 10^{18}$ cm$^{-3}$ and then decreases for higher doping. For $N_D=6.3 \times$ and 8$\times 10^{18}$ cm$^{-3}$, the combination of a strong smearing of the absorption edge and the Moss-Burstein shift reduces the index $\gamma$ to 0.64. Besides, the normalized reabsorption probability is redistributed to higher distances, leading to a broadened hole distributions in the layer  both at low and high ends of the concentration range. This correlates with the luminescence ratio dependence shown in Fig. \ref{TotIntenR}, namely with the observed minimum in luminescence intensity in transmission geometry  for samples with $N_D=2\times 10^{18}$ cm$^{-3}$.  We remark that the accuracy of the ${\cal P}(|z|)$ calculation is not high for heavily-doped samples due to the less reliable approximation for the emission spectrum.  

To calculate the stationary hole distribution, consider the temporary evolution of the distribution $p(z,t)$ after a short excitation pulse at $t=0$. Given the  $p(z,t)$, we can evaluate the stationary distribution $p_{st}(z)$ for constant excitation rate using the Duhamel principle,\cite{Duhamel} viz.
\be
p_{st}(z)=\int_0^\infty p(z,t)\exp(-t/\tau)dt/\tau~, \label{IntDuh}
\ee
where $\tau$ is the average lifetime of holes.
\begin{figure*}[t]
\begin{minipage}{0.47\linewidth}
\centering\epsfig{figure=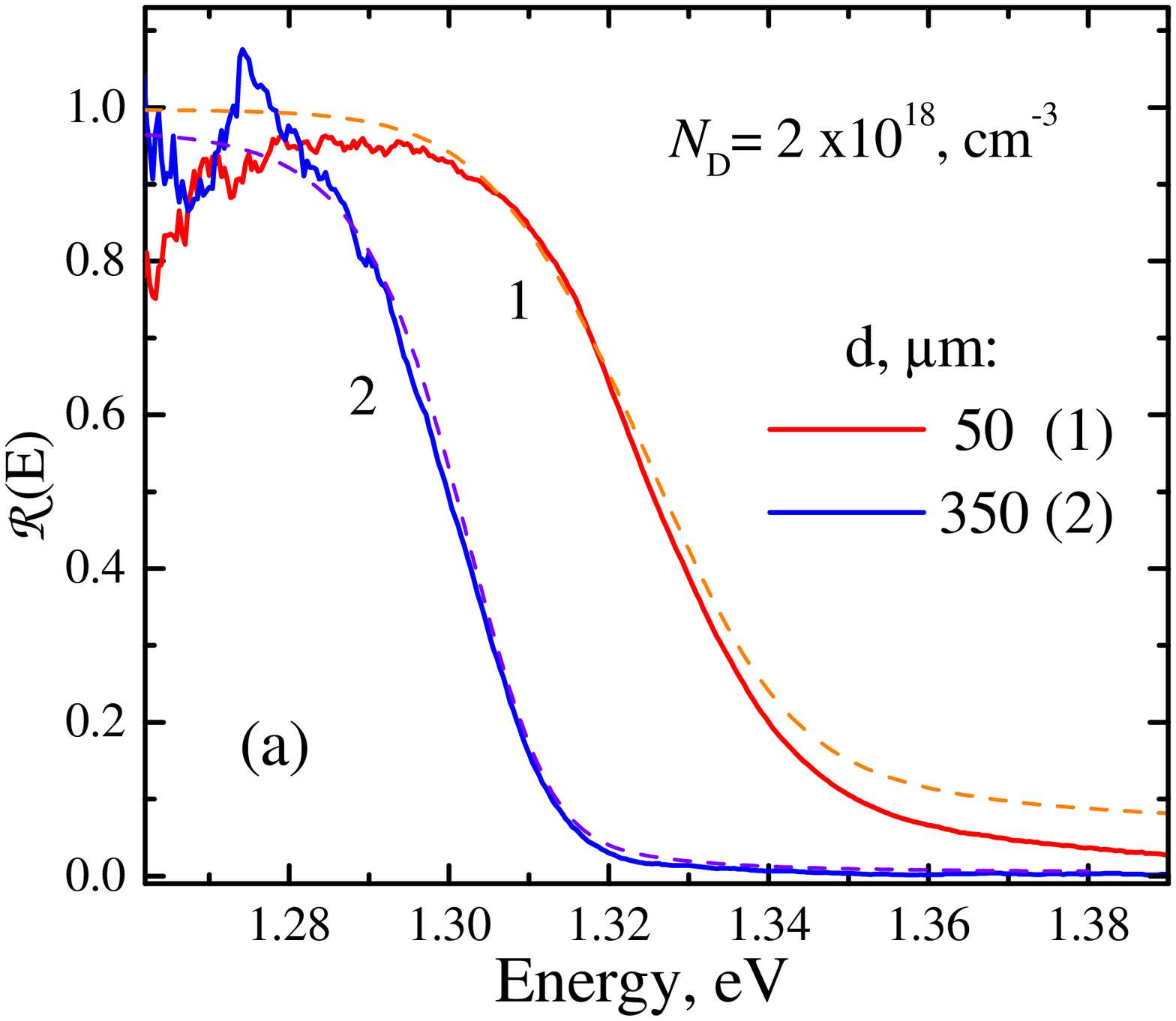,width=7.3cm}
\end{minipage}
\begin{minipage}{0.50\linewidth}
\centering\epsfig{figure=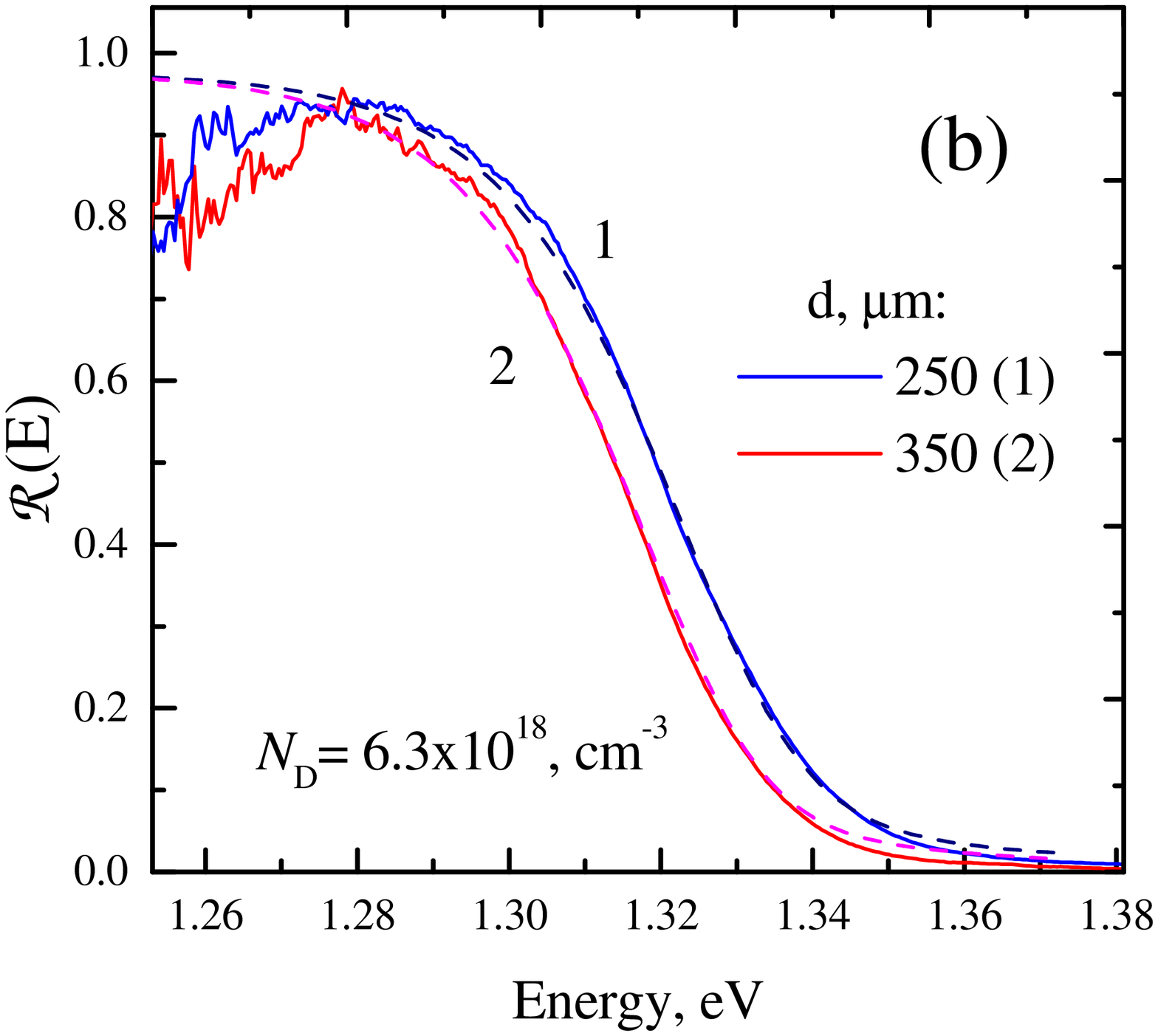,width=7.3cm}
\end{minipage} \hfill
\caption[]{(Color online) Spectral ratio $\mathcal{R}(E)$ for samples of different thickness $d$. (a) Results for moderately doped ($N_D=2 \times 10^{18}$ cm$^{-3}$) samples with $d=50$ and 350 $\mu$m (solid curves) are compared against model  calculations with  $\Phi = 32$ (dashed lines).  (b) Results (solid curves) for heavily doped ($N_D=6.3 \times 10^{18}$ cm$^{-3}$) samples with $d=250$ and 350 $\mu$m are fit by model calculations (dashed lines) with $\Phi = 10$. Back side is assumed to be non-reflecting. }
\label{Filt50Fit250}
\end{figure*}

Equation (\ref{IntDuh}) suggests an approach to the Monte Carlo calculation of $p_{st}(z)$.  First, we simulate the distribution $p(z,N)$ after a fixed number $N$ of recycling events of holes initially excited at the surface. This distribution can be viewed as a distribution $p(z,t)$ of the hole spread for a given time, $t=N \tau_r$, where $\tau_r$ is the radiative emission time.  

Next, we average the Monte-Carlo simulated distributions $p(z,t)$ over the random walk times $t$ distributed as $\exp(-t/\tau)/\tau$. This average is equivalent to averaging over the number of recycling events with the average recycling  factor $\Phi$,  
\be
p_{st}(z)=\sum_{N=1}^\infty p(z,N)\exp(-N/\Phi)~. \label{SumDuh}
\ee
For $\Phi \gg 1$  the main contribution to the sum at all $z > z_{min} $ comes from the terms with $N \gg 1$, so that Eqs. (\ref{SumDuh}) and (\ref{IntDuh}) give very close results. 


The simulated stationary hole distributions are sensitive to the boundary conditions. To discuss this effect, we consider the results obtained for a sample with  $N_D=6.3 \times 10^{18}$ cm$^{-3}$, see Fig. \ref{HoleDist} (a).  
The normalized distributions are found to be  practically insensitive to the boundary conditions on the front surface. Results presented in Fig.  \ref{HoleDist} are calculated with the reflective front (rf) boundary condition. In this case the stationary distributions satisfy $p_{rf}(z) \approx 2 p_{\infty}(z)$, where  $p_{\infty}(z)$ is the distribution for a localized source in an infinite medium.  This corresponds to the contribution  of an ``image'' source of photons provided by the reflecting boundary. 

In contrast, the distribution is substantially affected by the conditions at the back surface, see the curves 1-4 in Fig. \ref{HoleDist} (a). Depicted are the hole distributions for: reflective (r), absorbing (a), and ``non-reflective'' (nr) boundary conditions (the latter corresponds to a semi-infinite sample without a back surface). All distributions show a characteristic non-exponential drop-off with distance from the front surface, typical for L\'evy flights.  

In case of reflective back surface, the decline of concentration appeared to be much slower, $p_{rb}\propto z^{1+\gamma'}$ where $\gamma' \le 0$, as compared to that for the semi-infinite medium (or non-reflective back surface, curve 4). The slower declining distribution is in contradiction with the fast decline of $F_1(E)$ in the region of $\alpha$ above 200 cm$^{-1}$ and also with the drop-off of $\mathcal{R}(E)$ observed in the 50 $\mu$m sample.

For either the absorbing back surface or the non-reflective boundary condition, the distribution has power asymptotics $p(z)\propto z^{-1-\tilde \gamma}$ with $\tilde \gamma \approx 0.12 < \gamma$. It still extends to much wider region than an exponential distribution and decays slower than the reabsorption probability ${\cal P}(z)$. The assumption of non-reflective back boundary gives a much closer fit to out experimental data for the filtering functions in all samples. We have no physical reason to assume a totally light-absorbing back surface. But it seems reasonable that the surface scatters part of the luminescent radiation diffusively, so that it  escapes from the observation area and does not contribute to further recycling process. The luminescence escape from the observation area manifests itself also in the filtering function dependence in the transparency region at the red wing of the emission line.  Therefore, in our comparison with experimental data for the filtering functions and their ratios, we shall use the distributions obtained with ``non-reflective back boundary''.

The simulated distributions $p(z)$ for the set of differently doped samples, assuming non-reflective boundary conditions at the back surface, are shown in Fig. \ref{HoleDist} (b) in log-log scale.  For comparison, we also show an exponentially decaying distribution corresponding to $L_d=45$  $\mu$m (cf. Table \ref{tab:Ld}). We note that the non-exponential decay of the concentration is more sensitive to the recycling factor than to the variation of $\gamma$.

Using the hole distributions of Fig. \ref{HoleDist} and Eqs. (\ref{refFil}, \ref{trFil}), we calculated $\mathcal{R}(E)$ for all samples, using $\Phi$ as the only adjustable parameter. The best-fit values of $\Phi$  are also listed in Table \ref{tab:Kern}. These values are close to but somewhat smaller than the values $\Phi_l$  obtained independently from the time-resolved luminescence studies. \cite{Semyon1}
\begin{figure}[b]
\epsfig{figure=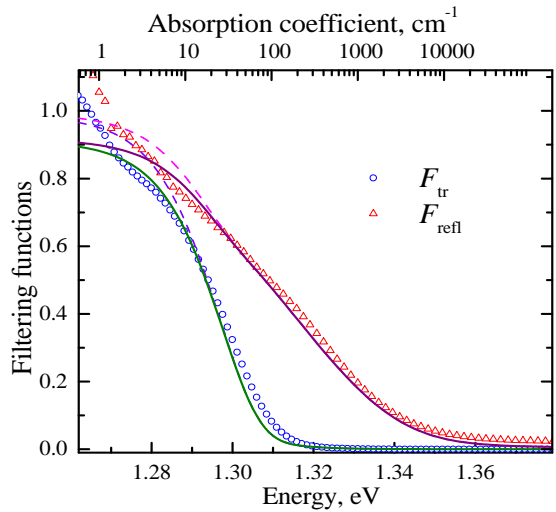,width=7.4cm,height=6.8cm} \caption[]
{(Color online) Filtering functions $F_{tr}(E)$ and $F_{refl}(E)$ (blue and red dots, respectively) for $N_D=3 \times 10^{17}$ cm$^{-3}$ sample and their comparison to model calculations. Hole distribution is obtained by Monte Carlo modeling with the reabsorption probability given by Eq. (\ref{probKer}) and the quantum efficiency $\eta =$ 98\%, ignoring multiple reflections;  dashed lines correspond to allowing multiple reflections.}
\label{Filt3e17}
\end{figure}
The most sensitive comparison of the theory and experiment is provided by the samples  with the same doping $N_D=2 \times 10^{18}$ cm$^{-3}$ but very different thickness, see Fig. \ref{Filt50Fit250} (a). Calculations with the modeled anomalous distributions give an excellent agreement with the experimental data. Even though for the relatively thick samples [$d$=250 and 350 $\mu$m, see Fig. \ref{Filt50Fit250} (b)] a satisfactory agreement can be also obtained assuming an exponential hole distribution, this assumption fails completely when one considers a pair of samples of vastly different thickness ($d$=50 and 350 $\mu$m). The exponential approximation fails to describe thin and thick samples simultaneously. 

The low-doped samples permit a more detailed comparison of the experimental and calculated spectra. If we assume that the initial spectra are well described by the van Roosbroeck-Shockley  relation (\ref{ShVR}), then we can use Eqs. (\ref{refFil},\ref{trFil}) to find  the filtering functions for the transmission and reflection spectra separately (as well as their ratio).   Comparison of the filtering functions for the sample with $n=3\times 10^{17}$ cm$^{-3}$ is shown in Fig. \ref{Filt3e17}. The results for the ratio do not differ noticeably from the experimental data shown in Fig. \ref{FiltRatio}. In the rapid decay region of the filtering functions, the agreement with experiment is quite good. However, there is a deviation in the region of small values of $\alpha(E)$. One should note that in this region the variation of functions $F_{tr}(E)$ and $F_{refl}(E)$ is sensitive to multiple reflections of the outgoing radiation from both the front and the back surfaces. Again, some of the angle-distributed radiation goes out of the observation area and is therefore lost. In the extreme case of a narrow observation area, only one-time reflection from the back surface contributes to the spectra. The filtering functions calculated barring multiple reflections are shown in Fig. \ref{Filt3e17} by the solid lines, while calculations allowing all multiple reflections are shown by the dashed lines. Since the experimental data in the red wing of the dependences $F_{tr}(E)$ and $F_{refl}(E)$ fall between these lines, we conclude that the discrepancy is likely due to a deviation of the experimental conditions from the assumed one-dimensional geometry. This, of course, does not invalidate the results in the wider energy range, where the filtering functions exhibit rapid variation. 

\section{discussion}
As was shown above, the experimental results for the filtering functions and their ratio are well 
interpreted in a model where the only mechanism of the hole spread in the sample is by photon recycling. Effects associated with the conventional mobility of holes reside in the parameters of the model, such as the recycling probability distribution. Thus,  the minimum recycling distance incorporates the hole diffusion length and the recycling factor $\Phi$ reflects the losses through all channels including surface recombination. 

The main result is that the hole concentration does not decay exponentially from the wafer surface, as it would for a normal diffusion process. The spread resulting from the photon-assisted random walk of holes is in agreement with the L\'evy-flights theory, which predicts heavy power-law tails in the distribution.

\begin{figure}[b]
\epsfig{figure=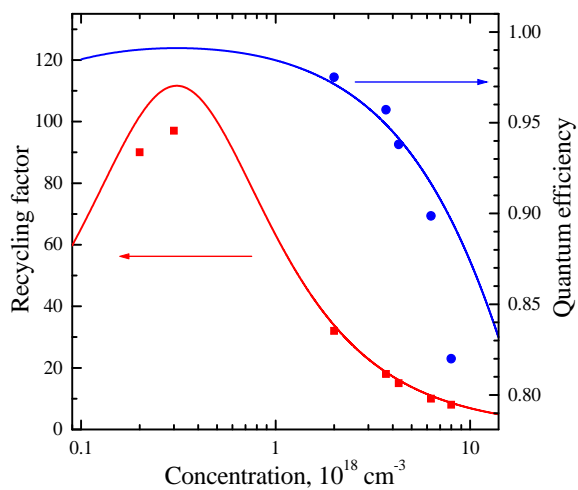,width=7.6cm,height=6.4cm} \caption[]
{(Color online) Quantum radiative efficiency $\eta$ (circles and solid line) from Ref. \onlinecite{Semyon1} and the recycling factor $\Phi$, both calculated from the earlier data (solid line) and obtained from the present experimental data (squares), cf. Table \ref{tab:Kern}.}
\label{QEandRec}
\end{figure}

It is interesting to compare the results for the quantum radiative efficiency $\eta$ and the corresponding recycling factor $\Phi$, obtained in this work, with those obtained earlier \cite{Semyon1} from the kinetics of time-resolved luminescence. The concentration dependence of these very important parameters is illustrated in Fig. \ref{QEandRec} for both the earlier and the present studies. Both sets of results are also listed in Table \ref{tab:Kern} in the instance of the recycling factor, as $\Phi_l$ and $\Phi$, respectively. The earlier data were obtained with the excitation photon energy in the relatively low-absorption part of the spectrum (red wing of the fundamental absorption edge) and therefore they correspond to a much larger penetration depth of the excitation and a different resulting hole distribution. Besides, in the earlier experiments \cite{Semyon1} the escape of photons from the observation spot on the sample surface was found to be of importance, providing and additional mechanism of the photon loss.

The data for the quantum radiative efficiency $\eta$ from Ref. \onlinecite{Semyon1} are shown by the circles in  Fig. \ref{QEandRec}. The solid line drawn through these circles is an interpolation based on the standard decomposition of hole recombination rates into the radiative ($B$) and non-radiative terms, impurity ($A$) and Auger ($C$), viz. $\eta (N_D) = BN_D/(A + BN_D + C N_D^2)$. The solid line for $\Phi$ shows the same data recalculated as $\Phi (N_D)=\eta/(1-\eta)$, which gives an upper estimate for the recycling factor. \cite{Fact} The squares show the results of the present work. Both the high values of the recycling factor and its concentration dependence are in close agreement with the earlier results, obtained by a very different technique. 

The most direct experimental manifestation of the non-exponential decay of the hole concentration is found in the ratio $\mathcal{R}(E)$ of the spectra observed in transmission and reflection geometries for the samples of the same doping level but very different thicknesses, see Fig. \ref{Filt50Fit250}. Indeed, the ratio $\mathcal{R}(E)$ will rapidly decline in energy only if the distribution itself decays within the thickness of the sample, as is supported by the data for  50 $\mu$m-thick samples, which suggest that the spread size is shorter than the sample thickness. This, however, is in contradiction with the fairly high intensity of  transmitted radiation for much thicker (350 $\mu$m) layers of the same doping. The power-law decay of the hole concentration is steep enough at short distances (steeper than an exponent) to fit the data for the thin sample, and at the same time slow enough at large distances (again, compared to an exponent) to account for the data for thick samples. The set of samples with low doping but different thicknesses should be most useful for further detailed investigations of the anomalous transport.

Our experimental data can be well described by the distributions obtained with different boundary conditions at the front (excitation) surface, since the concentration decay is relatively insensitive to these conditions. This is not the case for the back-surface boundary conditions. Our results are not in agreement with the assumption of almost total reflection from the back surface that one could expect based on the refractive index ratio. We attribute this discrepancy to the radiation escape from the observation region, in agreement with our earlier conclusions in time-resolved photoluminescence studies. \cite{Semyon1}   Good agreement is obtained for either absorbing or ``non-reflecting'' back surface.  

Our transport problem is well approximated by a statistical model that has been thoroughly studied, namely that of a one-dimensional random walk on an infinite line.\cite{Metz,Zol} Therefore, apart from some modifications at the back surface (see Fig. \ref{HoleDist}), a suitable distribution of holes after $N \gg 1$ flights originating from $z=0$  should approach the so-called standard {\em stable distribution}\cite{Zol} with zero asymmetry and a given index $\gamma$, 
\be
p(z,N) = \int_0^\infty \cos(k z)e^{-N(z_ck)^\gamma}d k~, \label{stable}
\ee where $z_c$ is the depth scaling factor. By the order of magnitude one has $z_c = z_{min}$ (comparison of the distribution (\ref{stable}) with our Monte-Carlo results for moderately doped samples gives $z_c = 0.23$ $\mu$m).  To find the stationary distribution for a given finite average number $\Phi$ of random flights, one should use Eq. (\ref{IntDuh}) and integrate Eq. (\ref{stable}), resulting in
\be
p(z,F) = \int_0^\infty \frac{\cos(k z)}{1+\Phi(z_ck)^\gamma} d k~. \label{stastbl}
\ee
Equation (\ref{stastbl}) yields a distribution that is close to our Monte Carlo results. Furthermore, it can be analyzed quantitatively. Analysis of Eq. (\ref{stastbl}) readily shows the existence of two regions in the excitation front spread, that of asymptotic decay and that of short flights. In the asymptotic region, the hole concentration is dominated by the one-pass long-distance flights and hence decays as the reabsorption probability  itself, $p(z) \propto z^{-\gamma-1}$. The asymptotic region corresponds to $z \gg   z_c \Phi^{1/\gamma} \equiv z_f$. For $\Phi \gg 1$ and $\gamma < 1$ the front spread distance $z_f \gg z_c$.  In the region $z\ll z_f$, large values of $k$ contribute to the integral in (\ref{stastbl}). Then one can neglect unity in the denominator of  (\ref{stastbl}), which gives $p(z)\propto z^{\gamma-1}$ with much slower decay with distance than in the asymptotic region. 

For moderately doped samples, one has  $z_f \approx 300$ $\mu$m and the onset of the asymptotic dependence is at distances larger than the sample thickness. Experimentally, neither $d\ll z_f$ nor $d\gg z_f$ conditions apply. This hampers the determination of the index of anomalous diffusion through the luminescence spectra. The experiments for the radiation  escape from the remote edge of the wafer \cite{SSCL} are better suited for this goal.

Finally, we note that the one-dimensional L\'evy flights transport is described by an integro-differential equation, corresponding to the diffusion equation with the recycling term.\cite{Rossin} This equation can be solved numerically, using available COMSOL software. Evaluation of the hole distribution in this way gave excellent agreement with the Monte Carlo results, except within an area of the order of the hole diffusion length near the sample surface. In the far-distance asymptotic region, the integro-differential equation admits of an analytic solution,\cite{Ivanov} which is again the distribution  (\ref{stastbl}).

\section{conclusion}
The luminescence spectra of moderately thick $n$-type InP wafers have been studied in reflection and transmission geometries. Analysis of the reflection and transmission spectra and especially of their ratios provides an unambiguous evidence for the anomalous non-diffusive transport of minority carriers via photon recycling. As a result, the minority-carrier distribution acquires heavy tails, 
accompanied by an anomalously high intensity of luminescence in the transmission geometry. Our interpretation of the experiment provides an independent estimate of the photon recycling factor $\Phi$, which is in good agreement with earlier results obtained by time-resolved luminescence studies.  

The analysis also confirms the very high  quantum radiation efficiency in moderately doped samples. Such samples provide, therefore, a remarkable experimental system for studying the L\'evy-flight transport. Its unique advantage is the possibility to vary the L\'evy distribution index by changing the temperature or the doping. 

\begin{acknowledgments}
This work was supported by the Domestic Nuclear Detection Office (DNDO) of the Department of Homeland Security, by the Defense Threat Reduction Agency (DTRA) through its basic research program, and by the New York State Office of Science, Technology and Academic Research (NYSTAR) through the Center for Advanced Sensor Technology (Sensor CAT) at Stony Brook. 
\end{acknowledgments}


\begin{thebibliography}{99}
\bibitem{cube} The factor $E^3$ rather than $E^2$ enters Eq. (\ref{ShVR}) since the spectrum we register describes the light power, not the number of photons.
\bibitem{Fact} Even though denoted by the same symbol $\Phi$, the average number of photon recycling events in an infinite space may differ from the recycling factor for a finite  layer excited at the surface.
\end{thebibliography}
\end{document}